\documentclass[aps,prl,twocolumn, groupedaddress]{revtex4-1}
\usepackage{graphicx}
\usepackage{amsmath}

\begin{document}


\title{Temperature and bias voltage dependence of Co/Pd multilayer-based magnetic tunnel junctions with perpendicular magnetic anisotropy}


\author{Zo\"e Kugler}
\email{zkugler@physik.uni-bielefeld.de}
\homepage{www.spinelectronics.de}
\author{Volker Drewello}
\author{Markus Sch\"afers}
\author{Jan Schmalhorst}
\author{G\"unter Reiss}
\author{Andy Thomas}
\affiliation{Bielefeld University, Department of Physics, Universit\"atsstr. 25, 33615 Bielefeld}
%
%
\date{\today}
\begin{abstract}
Temperature- and bias voltage-dependent transport measurements of magnetic tunnel junctions (MTJs) with perpendicularly magnetized Co/Pd electrodes are presented. Magnetization measurements of the Co/Pd multilayers are performed to characterize the electrodes. The effects of the Co layer thickness in the Co/Pd bilayers, the annealing temperature, the Co thickness at the MgO barrier interface, and the number of bilayers on the tunneling magneto resistance (TMR) effect are investigated. TMR-ratios of about 11\,\% at room temperature and 18.5\,\% at 13\,K are measured and two well-defined switching fields are observed. The results are compared to measurements of MTJs with Co-Fe-B electrodes and in-plane anisotropy.
\end{abstract}
%
%
\maketitle
%
\section{Introduction}
The tunneling magnetoresistance (TMR) effect in magnetic tunnel junctions (MTJs) is of interest for applications such as high-density read heads and non-volatile memory devices \cite{parkin1999}. A large TMR-ratio was theoretically predicted in 2001 for MTJs with a fully epitaxial (001) MgO barrier and (001) bcc Fe, Co, or Co-Fe as electrodes \cite{Butler2001,Mathon2001,Zhang2004}. The experimental realization of the large TMR effect is the basis for spin transfer torque (STT) switched high-density magnetic random access memory (MRAM) applications \cite{Parkin2004,Yuasa2004}. However, remaining challenges for planar MTJs are the increase of thermal stability for further miniaturization of the devices and the shape limitations due to magnetization curling at the edges of a patterned element \cite{zheng1997}.  MTJs based on magnetic layers with perpendicular magnetic anisotropy (PMA), so-called perpendicular magnetic tunnel junctions (pMTJs), are predicted to have a larger thermal stability as a result of a higher magnetic anisotropy and a lower switching current density for STT switching \cite{Mangin2006,Wang2006,nakayama2008,liu2009} compared to in-plane MTJs. In addition to the size limitations, the shape limitations of MTJ elements are also eliminated because pMTJs have no limit on the cell aspect ratio of patterned elements \cite{yoo2005,nishimura2002}.
Promising materials with perpendicular magnetic anisotropy are Co based multilayers, such as Co/Pd or Co/Pt \cite{carcia1985}. These materials have been successfully integrated into MTJs \cite{tadisina2010, Ye2008, park2008,law2007,kim2008,wang2010}.
In this study, temperature- and bias voltage-dependent transport measurements of MgO-based pMTJs with Co/Pd multilayers as electrodes are presented and compared to measurements of MTJs with Co$_{40}$Fe$_{40}$B$_{20}$ electrodes. The effects of the Co layer thickness in the Co/Pd bilayers, the annealing temperature, the Co thickness at the MgO barrier interface, and the number of Co/Pd bilayers on the TMR effect are investigated. Magnetic measurements of the Co/Pd multilayers are performed to characterize the electrodes of the tunnel junctions with respect to the Co layer thickness in the Co/Pd bilayers and to the annealing temperature.
%
\section{Preparation}
The samples are prepared in a magnetron sputter system with a base pressure of $1\times10^{-7}$\,mbar. The layer stacks are sputtered on top of a thermally oxidized (500\,nm) silicon (001) wafer. The layer stack of the samples is Si wafer\slash SiO$_2$\slash Ta\slash (Pd~1.8\slash Co~$t_{Co}$)$_X$\slash interface layer\slash MgO~2.1\slash  interface layer\slash(Co~$u_{Co}$\slash Pd~1.8)$_Y$\slash protection layers (all numbers in nm). The aim is to tailor different switching fields of the electrodes and a maximal TMR ratio. Therefore, the Co thickness $t_{Co}$ in the lower multilayer, the number of multilayers of the lower electrode $X$, the Co thickness in the upper multilayers $u_{Co}$, and the number of multilayers of the upper electrode $Y$ is changed. Additionally, the barrier interface is dusted with Co or Co$_{40}$Fe$_{40}$B$_{20}$.
The samples are annealed after sputtering at different temperatures for 60 minutes in a magnetic field of 6500\,Oe perpendicular to the film plane to enhance the perpendicular magnetic anisotropy of the Co/Pd multilayers and to crystallize the MgO barrier. The stack is patterned by laser lithography and ion beam etching. The resulting patterns are squares of $7.5\times7.5$\,$\mu$m$^2$ and $15.5\times15.5$\,$\mu$m$^2$. These structures are capped with gold contact pads.

The transport measurements are done by a conventional two-probe technique with a 10\,mV bias voltage in a perpendicular magnetic field. The low-temperature measurements are done in a closed-cycle helium cryostat (\textsc{Oxford} Cryodrive 1.5) with a temperature range of 13 to 330\,K.

\section{Results and Discussion}
%
%
%
%
%
\begin{figure}
\centering
	\includegraphics[]{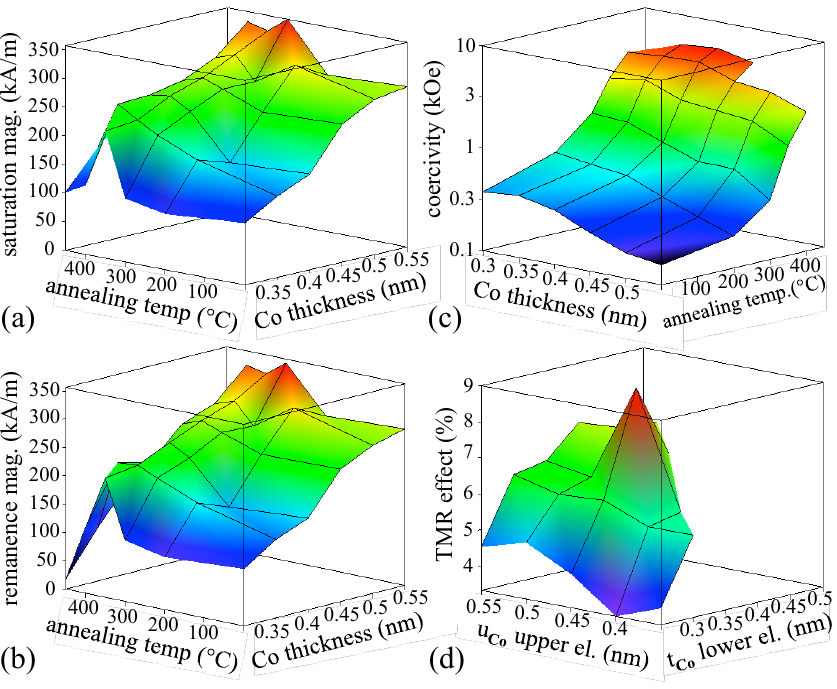}
	\caption{\label{3D} (a)-(c) Magnetic behavior of samples ...\slash (Pd~1.8\slash Co~$t_{Co}$)$_9$\slash MgO~2.1 (all numbers in nm) in dependence of the Co layer thickness $t_{Co}$ and annealing temperature $T_a$. (d) Room temperature TMR of samples ...\slash (Pd~1.8\slash Co~$t_{Co}$)$_9$\slash Co-Fe-B~1\slash MgO~2.1\slash  Co-Fe-B~1\slash (Co~$u_{Co}$\slash Pd~1.8)$_9$\slash ... as a function of the Co layer thickness $t_{Co}$ in the lower electrode and $u_{Co}$ in the upper electrode.}
\end{figure}
First, the magnetic behavior of the SiO$_2$\slash Ta\slash (Pd~1.8\slash Co~$t_{Co}$)$_X$\slash MgO~2.1electrodes with $X$=9 is investigated. The Co layer thickness  $t_{Co}$ and the annealing temperature $T_a$ are changed. The measurements are done with an alternating gradient magnetometer (\textsc{Micro Mag} 2900, \textsc{Princeton Measurements Corporation}). The magnetic field is applied perpendicular to the film plane. Figure \ref{3D}(a)-(c) shows the saturation magnetization ($M_s$), remanent magnetization (M$_r$), and coercivity ($H_c$) in dependence of $t_{Co}$ and of $T_a$. Here, $t_{Co}$ is changed from 0.3\,nm to 0.55\,nm in steps of 0.05\,nm. The annealing temperature is changed for every fixed $t_{Co}$ from room temperature (as-prepared state) to 450\,$^\circ$C. 
$M_s$ increases with increasing Co layer thickness, as shown in Figure \ref{3D}(a).
The annealing temperature dependence of $M_s$ shows that $M_s$ first increases with increasing $T_a$. The maximum $M_s$ is reached for annealing temperatures between 200\,$^\circ$C and 350\,$^\circ$C, dependent on the Co layer thickness. After reaching this maximum the $M_s$ decreases.
The lowest value of $M_s$ is about 108\,kA/m for 0.3\,nm Co in the as-prepared state, whereas the highest value of $M_s$ is 349\,kA/m at 0.55\,nm Co and 300\,$^\circ$C annealing temperature.

The $t_{Co}$ and $T_a$ dependence of $M_s$ and $M_r$ as well as their absolute values are very similar, as one can see by comparing Figure \ref{3D}(a) and \ref{3D}(b), which show $M_s$ and $M_r$, respectively. Thus, the squareness ($M_r$/$M_s$) is nearly one for all of the investigated samples. This shows the very strong PMA and good quality of the Co-based superlattices.

The coercivity, shown in Figure \ref{3D}(c), increases with the annealing temperature.
 After being roughly constant for Co thicknesses between 0.3\,nm and 0.4\,nm, dependent on the annealing temperature, the coercivity decreases with increasing $t_{Co}$.  The lowest value is about 157\,Oe for 0.55\,nm Co in the as-prepared state, whereas the highest value of $H_c$ is 5520\,Oe at 0.4\,nm Co and 450\,$^\circ$C annealing temperature. This coercivity values are in a typical range for Co/Pd multilayers \cite{Hashimoto1990} It is reasonable that the $H_c$ of the Co/Pd multilayers decreases with increasing Co thickness. For nine bilayers of Co/Pd, as used here, the critical Co layer thickness at which the perpendicular anisotropy turns into in-plane anisotropy is 1.15\,nm (not shown).

%
%
%
These magnetic measurements demonstrate, that it is possible to tune the switching field of the multilayers by changing the Co layer thickness in the Co/Pd multilayers. This is an important factor for the use of such multilayers as electrodes in pMTJs to achieve two well-defined switching fields and to stabilize the anti-parallel state. Consequently, the multilayers are integrated as electrodes in magnetic tunnel junctions. The resulting sample stack is SiO$_2$\slash Ta\slash (Pd~1.8\slash Co~$t_{Co}$)$_X$\slash interface layer\slash MgO~2.1\slash  interface layer\slash(Co~$u_{Co}$\slash Pd~1.8)$_Y$\slash protection layers. The Co layer thickness $t_{Co}$ of the lower electrode is changed from 0.25\,nm to 0.5\,nm to improve the TMR effect and the switching behavior of the pMTJs. For every sample with one fixed $t_{Co}$, the Co layer thickness $u_{Co}$ of the upper electrode is varied. The lower electrode is always used as the hard magnetic electrode and the upper electrode as the soft one. This means, that $u_{Co}$ is always larger than $t_{Co}$. For all samples, the numbers of bilayers $X$ and $Y$ in the electrodes is kept constant at $X$=$Y$=9, as already used for the AGM measurements. A 1\,nm layer of Co-Fe-B is used as an interface layer to improve the growth of the MgO barrier on top of the Pd/Co electrode \cite{mizunuma2009, Yakushiji2010}. All samples are annealed after sputtering at 150\,$^\circ$C.
Figure \ref{3D}(d) shows the TMR effect as a function of the Co layer thickness of the lower electrode ($t_{Co}$) and as a function of the Co layer thickness of the upper electrode ($u_{Co}$). A maximum TMR effect of about 9\,\% is observed for $t_{Co}$=0.35\,nm and $u_{Co}$=0.45\,nm. This maximum is in the middle of the investigated Co thicknesses: for smaller and larger $t_{Co}$ and $u_{Co}$, the TMR ratio decreases to about  3.3\,\%. The area-resistance products of the samples are in the range of 2 to 6\,M$\Omega \mu$m$^2$.

%
%
%
\begin{figure}
	\includegraphics{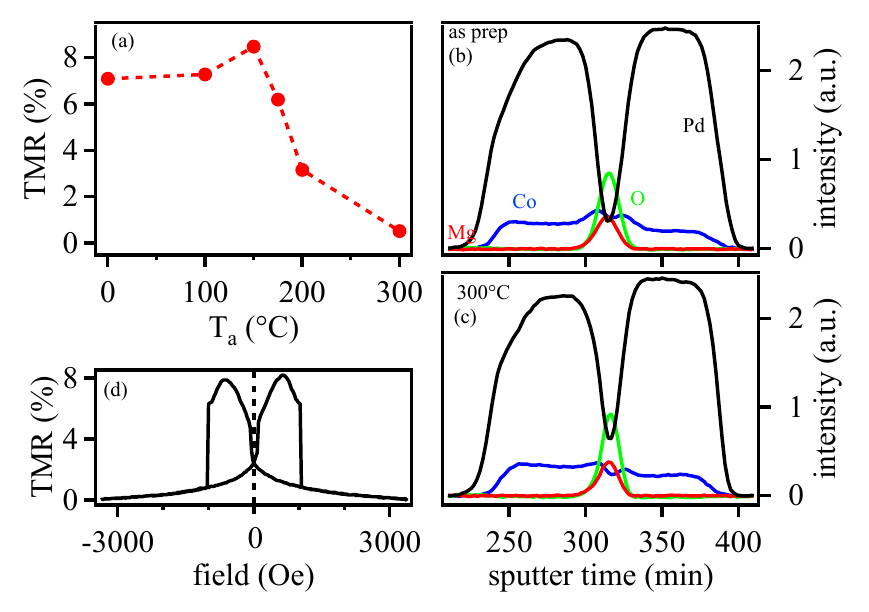}
	\caption{\label{TMRTa} (a) TMR as a function of annealing temperature for the sample stack ...\slash (Pd~1.8\slash Co~0.35)$_9$\slash Co-Fe-B~1\slash MgO~2.1\slash  Co-Fe-B~1\slash (Co~0.45\slash Pd~1.8)$_9$\slash ... (all numbers in nm). Composition depth profiles of Pd, Co, O, and Mg in the sample (b) before and (c) after annealing at 300\,$^\circ$C for 1 hour. (d) Major loop of the sample annealed at 150\,$^\circ$C.}
\end{figure}
Next, the annealing temperature $T_a$ is optimized. Figure \ref{TMRTa}(a) shows the $T_a$ dependence of the TMR effect for the  best sample from Figure \ref{3D}(d), with $t_{Co}$=0.35\,nm, and $u_{Co}$=0.45\,nm, $X$=$Y$=9 and 1\,nm Co-Fe-B as the interface layer. The TMR increases from about 7\,\% in the as-prepared state to about 9\,\% at an annealing temperature of 150\,$^\circ$C. For $T_a$ larger than 150\,$^\circ$C, the TMR decreases to 0.6\,\% at 300\,$^\circ$C. This strong decrease of the TMR for higher annealing temperatures is due to Pd diffusion to the MgO barrier. In Figure \ref{TMRTa}(b) and \ref{TMRTa}(c), the composition depth profiles of Pd, Co, O, and Mg are shown for the sample in the as-prepared state and after annealing at 300\,$^\circ$C for one hour. The profiles are measured by auger electron spectroscopy (AES) depth profiling.
The Pd intensity in the region of the MgO is higher in the measurement of the annealed sample, whereas a comparison of the Co intensities in the MgO region shows less Co after annealing. This indicates a diffusion of Pd to the MgO interface, where the Pd replaces the Co at high $T_a$, thereby reducing the TMR effect.
Figure \ref{TMRTa}(d) shows the major loop of the sample annealed at the optimum temperature of 150\,$^\circ$C for one hour. The increase of the TMR before reaching zero field is caused by an additional in plane component which is most likely produced by the Co-Fe-B layers.
%
%
%
To prove this assumption, Co-Fe-B was replaced by a Co interface layer with different thicknesses $v$.
\begin{figure}
	\includegraphics{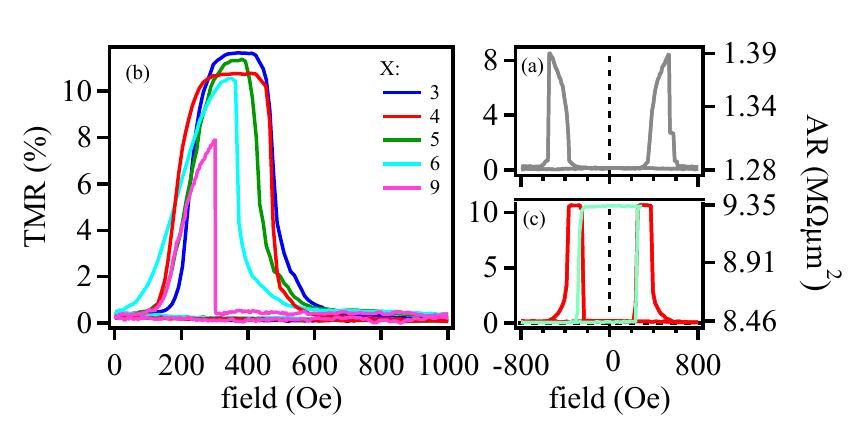}
	\caption{\label{TMRN} (a) Major loop of sample ...\slash (Pd~1.8\slash Co~0.35)$_9$\slash Co~0.7\slash MgO~2.1\slash  Co~0.7\slash (Co~0.45\slash Pd~1.8)$_9$\slash .... (all numbers in nm) (b) TMR loops of samples ...\slash (Co~0.6\slash Pd~1.8)$_X$\slash Co~0.7\slash Mg~0.5\slash MgO~2.1\slash Co~0.7\slash (Co~0.7\slash Pd~1.8)$_2$\slash ... with different numbers X of Co/Pd bilayers in the lower electrode. (c) Major and minor loop of the sample with X=4. }
\end{figure}
Figure \ref{TMRN}(a) shows the major loop of the sample with $v$=0.7\,nm. This sample shows the highest TMR ratio of about 9\,\%, similar to the TMR ratio of the sample with 1\,nm of Co-Fe-B at the MgO interface, but no in-plane component can be concluded from the TMR measurement. The hard electrode has a sharp switching field, whereas the magnetization of the soft electrode turns slowly. This is likely due to the strong magneto-static interaction between the soft and hard magnetic layers in the patterned pMTJs. The stray field of one of the perpendicular electrodes on the other electrode always assists the parallel alignment of the magnetization.

The number $X$ of Co/Pd bilayers in the lower electrode is changed from 9 to 3 to control the magneto-static interaction and to ensure the anti-parallel alignment of the magnetization in the electrodes over a certain field range. Figure \ref{TMRN}(b) shows the major loops in the positive field range for 0.7\,nm of Co. The upper electrode is kept constant at $Y$=2 with $u_{Co}$=0.7\,nm and $t_{Co}$ as 0.6\,nm. A thin Mg layer was inserted under the MgO to enable better growing of the barrier. The samples show a TMR effect of about 11\,\% at room temperature and an area resistance product of about 8.5\,M$\Omega \mu$m$^2$. The switching field of the hard electrode increases from 300\,Oe for $X$=9 to about 640\,Oe for $X$=3. Whereas the magnetization of the soft electrode turns slowly and has no sharp switching field for $X$=9, a hard switching is reached for $X$$<$5. 
The sample with $X$=4 shows two well-defined switching fields, and the minor loop has two separated magnetic states at zero field as shown in Figure \ref{TMRN}(c).

%
%
%
\begin{figure}
	\includegraphics{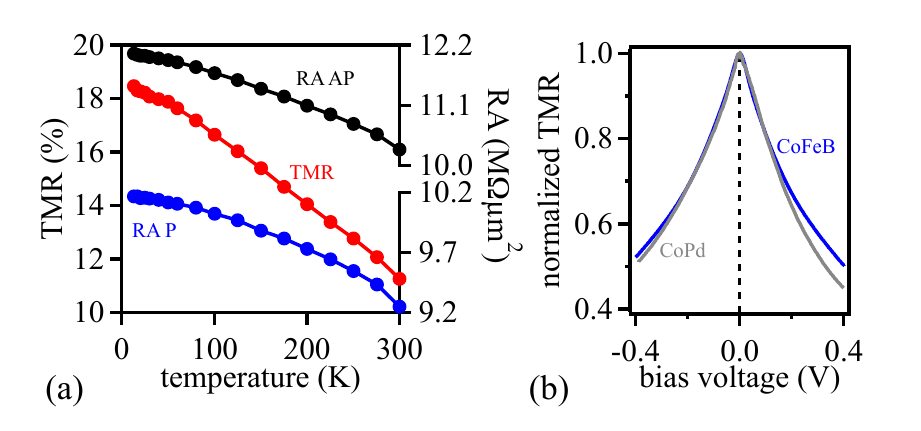}
	\caption{\label{TMRT}(a) Temperature dependent transport measurement of the sample ...\slash (Co~0.6\slash Pd~1.8)$_4$\slash Co~0.7\slash Mg~0.5\slash MgO~2.1\slash Co~0.7 \slash (Co~0.7\slash Pd~1.8)$_2$\slash ... (all numbers in nm). (b) Normalized TMR as a function of the bias voltage for the pMTJ and for a reference sample ...\slash Mn-Ir 12\slash Co-Fe-B~4\slash Al~1.2+Oxidation\slash Co-Fe-B~4\slash Ni-Fe~3\slash ... with in-plane anisotropy measured at 13\,K.}
\end{figure}
Finally, transport measurements for the optimized sample annealed at 150\,$^\circ$C were performed at different temperatures.  Figure \ref{TMRT}(a) shows the temperature dependence of the TMR and the area resistance products in the parallel state P and in the anti-parallel state AP. The area resistance product in the P state changes from 9.2 to 10.2\,M$\Omega\mu$m$^2$ with the temperature. This is a change of 9.8\,\%, which is in the typical range for the P state for MTJs \cite{drewello2008, Schmalhorst2007}. The area resistance product in the AP state changes by 14.5\,\%, which changes the TMR overall by a factor of 1.7 to 18.5\,\% at 13\,K. This is the highest value reported for pMTJs with Co/Pd multilayers as perpendicular magnetized electrodes \cite{mizunuma2009, lim2005}.
In Figure \ref{TMRT}(b), the bias voltage dependence of the TMR effect is shown and compared to an reference sample with Co-Fe-B electrodes and in-plane anisotropy. 
The MgO barrier of the pMTjs is not considered to be textured, as the TMR-effects are not close to that in the coherence tunneling scheme in epitaxially MgO-bcc systems. Furthermore, x-ray diffraction measurements show fcc (111) orientation of the Co/Pd electrodes (not shown). Even for MgO barriers with (111) texture coherent tunneling can not be expected \cite{hauch2008}. Thus, a MTJ with an amorphous Alumina-based barrier is chosen as the reference sample.
The reference sample is annealed at 275$^\circ$C and shows a TMR effect of 110\,\% at 10\,mV and 13\,K.  The bias voltage dependence of the TMR for the pMTJ with Co/Pd electrodes shows no significant change compared to the in-plane reference sample. A peak-like maximum is observed for zero bias, and the TMR decreases to the half-value for bias voltages of about 400\,mV.
\section{Summary}
In summary, temperature- and bias voltage-dependent transport measurements of pMTJs with Co/Pd electrodes were shown. The temperature- and bias voltage-dependent behaviors of pMTJs show no significant changes compared to MTJs with Co-Fe-B electrodes and in-plane anisotropy.
Magnetic measurements of the Co/Pd multilayers were performed to characterize the electrodes of the tunnel junctions. The Co/Pd multilayer films show strong PMA, even in the as-prepared state. The effects of the Co layer thickness in the multilayer electrodes, the annealing temperature, the Co thickness at the MgO barrier interface, and the number of Co/Pd bilayers on the TMR effects were investigated. A TMR effect of about 11\,\% at room temperature and 18.5\,\% at 13\,K was measured for an optimum annealing temperature of 150\,$^\circ$C.

\end{document}